\documentclass[12pt]{article}

\usepackage{graphicx}

\begin{document}

\newcommand{\Om}{\Omega}
\newcommand{\df}{\stackrel{\rm def}{=}}
\newcommand{\co}{{\scriptstyle \circ}}
\newcommand{\de}{\delta}
\newcommand{\lb}{\lbrack}
\newcommand{\rb}{\rbrack}
\newcommand{\rn}[1]{\romannumeral #1}
\newcommand{\msc}[1]{\mbox{\scriptsize #1}}
\newcommand{\dsp}{\displaystyle}
\newcommand{\scs}[1]{{\scriptstyle #1}}

\newcommand{\ket}[1]{| #1 \rangle}
\newcommand{\bra}[1]{| #1 \langle}
\newcommand{\vac}{| \mbox{vac} \rangle }

\newcommand{\e}{\mbox{{\bf e}}}
\newcommand{\va}{\mbox{{\bf a}}}
\newcommand{\bc}{\mbox{{\bf C}}}
\newcommand{\br}{\mbox{{\bf R}}}
\newcommand{\bz}{\mbox{{\bf Z}}}
\newcommand{\bq}{\mbox{{\bf Q}}}
\newcommand{\bn}{\mbox{{\bf N}}}
\newcommand {\eqn}[1]{(\ref{#1})}

\newcommand{\cp}{\mbox{{\bf P}}^1}
\newcommand{\n}{\mbox{{\bf n}}}
\newcommand{\sbz}{\msc{{\bf Z}}}
\newcommand{\sn}{\msc{{\bf n}}}

\newcommand{\be}{\begin{equation}}\newcommand{\ee}{\end{equation}}
\newcommand{\bea}{\begin{eqnarray}} \newcommand{\eea}{\end{eqnarray}}
\newcommand{\ba}[1]{\begin{array}{#1}} \newcommand{\ea}{\end{array}}

\newcommand{\cleqn}{\setcounter{equation}{0}}
\makeatletter
\@addtoreset{equation}{section}
\def\theequation{\thesection.\arabic{equation}}
\makeatother

\def\npb{Nucl. Phys. {\bf B}}
\def\plb{Phys. Lett. {\bf B}}
\def\mpla{Mod. Phys. {\bf A}}
\def\ijmpa{Intern. J. Mod. Phys. {\bf A}}
\def\cmp{Comm. Math. Phys.}
\def\prd{Phys. Rev. {\bf D}}

\def\vu{\vec u}
\def\vs{\vec s}
\def\vv{\vec v}
\def\vt{\vec t}
\def\vn{\vec n}
\def\ve{\vec e}
\def\vp{\vec p}
\def\vk{\vec k}
\def\vx{\vec x}
\def\vz{\vec z}
\def\vy{\vec y}
\def\vxi{\vec\xi}


\begin{flushright}
La Plata Th-05/01\\December 2005
\end{flushright}

\bigskip

\begin{center}

{\Large\bf
The fundamental non critical string
}
\footnote{This work was partially supported by CONICET, Argentina
}
\bigskip
\bigskip

{\it \large Adri\'{a}n R. Lugo and Mauricio B. Sturla}
\footnote{ {\sf
lugo@fisica.unlp.edu.ar, $\;\;$ sturla@fisica.unlp.edu.ar} }
\bigskip

{\it Departamento de F\'\i sica, Facultad de Ciencias Exactas\\
Universidad Nacional de La Plata\\ C.C. 67, (1900) La Plata, Argentina
}
\bigskip
\bigskip

\end{center}

\begin{abstract}
We obtain the (super) gravity solution in arbitrary space-time
dimension less than ten, that gives a low energy description of a
fundamental string embedded in a non critical vacuum, product of
$d$-dimensional Minkowski space-time and a cigar-like geometry
with scale $r_0$. This solution, one of the few known examples of
objects doubly localized, both at the origin of the transverse
space as well as at the tip of the cigar, is determined by its
charge $Q$ under the Kalb-Ramond gauge field $B$, and presumably
preserves, for even $d$, $2^\frac{d}{2}$ supercharges. Moreover,
we show that the solution is reliable at least in a region far
away from both origins, as it is the case with the well known
branes of critical string theory.

\end{abstract}
\bigskip

\section{Introduction}
\cleqn
Non critical string theories (NCST) saw the light in the early eighties, when A. Polyakov
\cite{pol} showed that it might be possible to define a theory of strings moving in dimensions
lesser than the so called critical dimension $D=26$ ($D=10$ in the superstring),
if we take into account an ``extra" degree of freedom or dimension, with dynamics governed
by the Liouville theory.
Much work was made on the subject, in particular in $D\leq 2$ and the related matrix
models \cite{ginmoo}.
However, the so called ``$c=1$ barrier"  seemed to forbid to go to higher
dimensions \cite{sei}.
Things change dramatically with the introduction of $({\cal N}_L,{\cal N}_R)=(2,2)$
superconformal symmetry on the world-sheet.
In reference \cite{kutasei}, it was showed that non critical superstring theories can be
formulated in $d=2n , n=0,1,\dots,4$ space-time dimensions, and describe consistent
solutions of string theory in sub-critical dimensions, with space-time supersymmetry
consisting of, at least, $2^{n+1}$ supercharges.
On the world-sheet, these theories present, in addition to the dynamical Liouville mode $\phi$,
a compact boson $X$, being the target space of the general form
\be
{\cal M}^d\times \Re_\phi\times S^1 \times M/\Gamma\qquad ,
\ee
where ${\cal M}^d$ is $d$-dimensional Minkowski space-time, $M$ is an arbitrary
$(2,2)$ superconformal field theory, and $\Gamma$ a discrete subgroup acting on
$S^1 \times M$.
Now, the Liouville theory include a linear dilaton background that yields a strong
coupling singularity; however the existence of the compact boson allow to resolve this
singularity by replacing the $\Re_\phi\times S^1$ part of the background by the
${\cal N}=2$ Kazama-Suzuki supercoset $SL(2,\Re)_k/U(1)$.
This space has a two-dimensional cigar-shaped geometry, with a natural scale given by
$r_0 = \sqrt{k\,\alpha'}$.
It provides a geometric cut-off for the strong coupling singularity, while coinciding with
the linear dilaton solution in the weak coupling region.
We will call this sector of the theory as ``the cigar".
In particular we will be interested in the vacuum solution (no $M$)
\be
{\cal M}^d\times SL(2,\Re)_k/U(1)= {\cal M}^d\times {\it Cigar}\label{vac}
\ee
From the world-sheet conformal field theory point of view, this theory has a
central charge,
\be
c = \frac{3}{2}\,d + 3\,\frac{k+2}{k} + \frac{3}{2} - \frac{3}{2} = \frac{3}{2}\,d
+ 3 + \frac{6}{k}\qquad .
\ee
Asking for the cancellation against the ghost contribution $c_{gh}= -15$, fixes
the level of the coset model to be,
\be
k= \frac{4}{8-d}\qquad .
\ee
It is known that this vacuum solution is an exact solution to all orders
in type II theories, up to a trivial shift $k\rightarrow k-2$ \cite{bs}.
In the $d=8$ case ($k=\infty$) we go back to the critical superstring in flat ten
dimensional Minkowski space-time.

It is natural to ask, as in the case of critical string theories (CST) \cite{argu},
for the solution of the (non critical) string vacuum equations of motion that
represents a string sited at the origin of the transverse $\Re^{d-2}$ flat space,
and at the tip of the cigar.
This is the question that we address in this paper.

A very important point is that the presence of a fundamental string (or more generally,
a set of p-branes) will not yield, in general, a solution that can be represented as
an exact conformal field theory, and the solutions will present higher order corrections.
What is more, as it is mentioned in references \cite{km}, \cite{ks1}, \cite{ks2},
where the authors focused on solution of the $AdS$ type, the presence of the cosmological
constant term in the non critical case could be enough dangerous to invalidate the
supergravity approximation in any case.
We will see however, that the solution obtained is trustworthy in a (limited) region of
the space, in much the same way that the well known brane solutions are in the critical
case \cite{hs}.
\bigskip

\section{The low energy effective field theory}
\cleqn

Let us consider the fields that represent the massless modes of a string moving on a
$D$ dimensional manifold $M_D$; they are the metric $G_{mn}$, the dilaton $\Phi$,
and the antisymmetric tensor or Kalb-Ramond field $B_{mn}$.
The bosonic part of the effective (super) gravity field theory that
describes their low energy dynamics is given by \cite{cfmp}, \cite{km},
\begin{eqnarray}
S[G,B,\Phi]&=& \int \epsilon_G\; e^{-2\Phi}\left(R[G]+4\left(D\Phi\right)^2\right.\nonumber
+\left.\Lambda^2-{1\over 2}\; H^2 \right)\label{acciongra}
\end{eqnarray}
Here $H=dB$ is the three-form field strength, $H^2 \equiv \frac{1}{2}\, H_{mn}\,H^{mn}$,
and $\epsilon_G=d^D x\,\sqrt{-\det G}\, $ is the volume form of the $D$ dimensional space-time.
\footnote{
We have not included tachyons or Ramond-Ramond fields in (\ref{acciongra}), because they
play no role in the solutions we are going to look for.
Furthermore, unless otherwise stated, we will be working in the string frame.
}
The cosmological constant is $\Lambda^2 = \frac{2 (26-D)}{3\alpha'}$ in the bosonic
string, and $\Lambda^2 = \frac{10- D}{\alpha'}=\frac{4}{r_0{}^2}\,$
in the superstring, and it is absent in the critical dimension.

On the other hand, let us consider a fundamental string with embedding defined by
${\bar X ^m (\sigma)}$ , where $\sigma\equiv (\sigma^0 ,\sigma^1 )$ are coordinates on
its world-sheet $\Sigma$.
Such a fundamental string couples with charge $Q$ to the $B$-field according to the
source term
\be
S_{F_1}[\bar X; B] =  Q\;\int_{\Sigma}\; B|_{pull-back} = \int_{M_D}B\wedge *J\;\;\;\;,
\ee
where the two-form current is defined by
\footnote{
The covariant delta function is defined by
\be
\delta^D_G(x-\bar{X})\equiv{1\over\sqrt{-\det G}}\;\delta^D(x-\bar{X})
\;\;\;\;\;,\;\;\;\;\; \int_{M_D}\;\epsilon_G\; \delta^D_G(x-\bar{X}) = 1\;\;\;,
\ee
and indices are lowered and raised with the metric $G$.}
\bea
J &=& \frac{1}{2}\, J_{mn}(x;\bar X)\, dx^m\wedge dx^n\cr
J^{mn}(x;\bar{X}) &\equiv&  Q\;\int_{\Sigma}d\bar{X}^m\wedge d\bar{X}^n\;
\delta^D_G (x-\bar{X}(\sigma))\;\;\;\;,\label{current}
\eea
From the action
\be
S = S[G,B,\Phi] + S_{F_1}[\bar X; B]\;\;\;\;,
\ee
the equations of motion that follow are,
\begin{eqnarray}
R_{mn}&=&{1\over2}\, H_{mp}\,H_n{}^p - 2D_m D_n\Phi\cr
\Lambda^2&=&e^{2\Phi}\; D^2(e^{-2\Phi})- H^2\cr
d\left(e^{-2\Phi}*H\right)&=& -*J\label{eqnmot}
\end{eqnarray}

In absence of sources, these equations of motion admit as solution the direct product
of $d$-dimensional Minkowski space-time and the cigar, i.e. the space (\ref{vac}),
\bea
G_0 &=& \eta_{1,d-1} + \tilde g\cr
\Phi_0(\tilde\rho) &=& \tilde\Phi_0 + \tilde\Phi(\tilde\rho)\cr
H_0 &=& 0\;\;\;\;\;\;\;\;\;\;\;\qquad,\label{snvac}
\eea
where $\,(\tilde g ,\tilde\Phi)$ is the cigar solution
($\tilde\rho\in\Re^+\,,\, \tilde\theta\in S^1$),
\bea
\tilde g &=& r_0{}^2 \; \left( d^2\tilde\rho + \tanh ^2 \tilde\rho\; d^2\tilde\theta
\right)\cr
e^{-\tilde\Phi(\tilde\rho)} &=& \cosh\tilde\rho\qquad.\label{sncigar}
\eea
Here $\tilde\Phi_0$ is the value of the dilaton at the tip of the cigar $\tilde\rho =0$,
and $r_0=\sqrt{k\,\alpha'}$ is the scale of the curvature,
$R[\tilde g]= \frac{4}{r_0{}^2}\, (\cosh\tilde\rho)^{-2}$.
We remark at this point that the effective string coupling constant,
$g_s \equiv e^{\Phi_0(\tilde\rho)}$, is bounded at $\tilde\rho =0$ by $e^{\tilde\Phi_0}$, that is a free
parameter of the theory.

Let us note for future use that the cigar is conformal to two dimensional flat space
(with cartesian coordinates $\vec z$) in the way,
\be
\delta \equiv d^2 z^1 + d^2 z^2 = e^{-2\,\tilde\Phi(\tilde\rho) }\;
\tilde g = d^2\rho + \rho^2\; d^2\theta \;\;\;\;,\;\;\;\;
\left\{\begin{array}{rcl}
\rho &=& r_0\,\sinh\tilde\rho\cr\theta &=& \tilde\theta
\end{array}\right.\label{flat2dmetric}
\ee

\section{The fundamental string}
\cleqn

Let us take the background (\ref{snvac}), (\ref{sncigar}) as the vacuum, and
let us consider a B-charged, one dimensional object  along $(x^0,x^1)\,$ in ${\cal M}^d$,
and localized at the origin $r\equiv|\vec y|=0$ of the orthogonal $\Re^{d-2}$
(with cartesian coordinates $\vec y$), and at the tip of the cigar $\tilde\rho = 0$.

An ansatz for the solution that represents the long range fields generated
by this long string, and that respects invariance under Poincar\'e, $SO(d-2)$ and $SO(2)$
transformations in the world-sheet of the string, $\Re^{d-2}$ and cigar respectively,
is the following one,
\bea
G &=& A^2(r;\tilde\rho)\; \eta_{1,1} + B^2(r;\tilde\rho)\; (dr^2+ r^2\; d^2\Omega_{d-3} )
+ C^2(r;\tilde\rho)\; \tilde g \cr
\Phi &=& \Phi(r; \tilde\rho)\cr
B&=& \left( E(r;\tilde\rho) - 1\right)\, dx^0\wedge dx^1
\;\;\longrightarrow\;\; H = dx^0\wedge dx^1 \wedge dE\qquad.\label{ansatz}
\eea
By using this ansatz, the current (\ref{current}) is given by
\be
J= -Q\,\delta^{d-2}_\delta (\vec y)\;\delta^2_{\delta}(\vec z)\; \frac{
e^{-2\tilde\Phi(\tilde\rho)}\,A^2}{B^{d-2}\,C^2}\;dx^0 \wedge dx^1\qquad ,
\ee
and we can write the equations (\ref{eqnmot}) in terms of our unknown functions
$(A, B, C, \Phi, E)$.
Instead of writing down the general equations, we do so after making the following
assumptions,
\be
B=C=1 \qquad,\qquad e^{2\Phi}=e^{2\Phi_0}\, A^2\qquad,\qquad
\epsilon\, E = A^2 - 1\qquad.\label{assump}
\ee
where $\epsilon^2 = 1$.
They leave us with just one unknown, the function $A$, and a sign.
If we further define $U \equiv A^{-2}$, and we use the vacuum equations
\footnote{
We denote with ``$\tilde{~}$", the variables, derivatives, etc., on the cigar.
}
\bea
\tilde{R}_{\tilde a\tilde b} &=&-2\tilde{D}_{\tilde a}\tilde{D}_{\tilde b}\tilde{\Phi}\cr
\Lambda^2&=& e^{2\tilde\Phi}\; \tilde D^2(e^{-2\tilde\Phi})\;\;\;\;,
\eea
it is possible to show, with the help of the relations,
\bea
*d(e^{-2\Phi}*H) &=&-A^4\, dx^0\wedge dx^1 \, D^m\left({e^{-2\Phi}\over
A^4}D_m(E)\right)\cr
e^{2\Phi} D^m\left( e^{-2\Phi}\,D_m (U)\right) &=&
e^{2\tilde\Phi} D^m_{(0)}\left( e^{-2\tilde\Phi}\,{D_{(0)}}_m(U)\right)\qquad ,
\label{relations}
\eea
that (\ref{eqnmot}) reduces to the differential equation
\be
- e^{2\tilde{\Phi}}D^m_{(0)}(e^{-2\tilde{\Phi}}D_{(0)m}(U)) =
\epsilon\, Q\,e^{2\tilde\Phi_0}\, \delta^{d-2}_\delta(\vec y)\; \delta^2_\delta(\vec z)\label{eqnU}
\ee
In equations (\ref{relations}) and (\ref{eqnU}), the suffix $``_{(0)}"$ refers to
covariant derivatives w.r.t. the vacuum metric (\ref{snvac}); explicitly,
\be
l.h.s.\,(\ref{eqnU}) =  -{1\over r^{d-3}}(r^{d-3}U^{'})^{'}-
e^{2\tilde{\Phi}} \tilde{D}_{\tilde c}\left( e^{-2\tilde{\Phi}}\tilde{D}_{\tilde c}(U)\right)
\equiv \left( \Delta + \tilde{L}_0 \right)(U)\qquad .
\ee
We recognize in the first term the laplacian operator in flat $(d-2)$ space, while that in
the second term, we introduced the operator
\be
\tilde{L}_0 \equiv -e^{2\tilde{\Phi}}\;\tilde{D}_{\tilde c}\;e^{-2\tilde{\Phi}}\;
\tilde{D}_{\tilde c}= -\frac{1}{r_0{}^2}\;\left( \partial^2_{\tilde\rho} +
2\,\coth(2\tilde\rho)\;\partial_{\tilde\rho} + \coth^2\tilde\rho\;\partial^2_{\tilde\theta}
\right)\qquad .\label{L0}
\ee
Resuming, the function $U$ is determined by the differential equation,
\be
\left( \Delta + \tilde{L}_0 \right)(U) =
\epsilon\,Q\,e^{2\tilde\Phi_0}\, \delta^{d-2}_\delta(\vec y)\; \delta^2_\delta(\vec z)\label{eqnUbis}
\ee
It is worth to note that the delta-sources appearing at the r.h.s. of the equations
(\ref{eqnU}) (\ref{eqnUbis}) are w.r.t. {\it flat} metrics.
The solution to this equation has, for $d>2$, the integral representation (see the Appendix
for details about its construction),
\bea
U(r,\rho) &=&1 + \epsilon\,Q\, e^{2\tilde\Phi_0}\, a_d\,
\int_0^\infty d\alpha \; \alpha^{-\frac{d-2}{2}} \;e^{-\alpha - {r^2\over 4r_0^2\alpha}}\;
I(\alpha ; \rho)\cr
I(\alpha;\rho)&=& \int_0^\infty ds \;\rho_0(s)\; e^{-4\alpha s^2}
F\left(\frac{1}{2} + is, \frac{1}{2}-is;1;-{\rho^2\over r_0^2}\right)\label{U}\cr
a_d{}^{-1}& =& 2^{d-3}\pi^{d+2\over 2} r_0{}^{d-2}\qquad .
\eea
Recapitulating, we have found a configuration consisting of a one dimensional,
$B$-charged object embedded in the vacuum space (\ref{vac}), localized at the origin
of the transverse $\Re^{d-2}$, and at the tip of the cigar, that we naturally identify as
the fundamental non critical string.
Such a background is given by
\bea
G &=& U^{-1}\; \eta_{1,1} + dr^2+ r^2\; d^2\Omega_{d-3} + \tilde g \cr
e^{2\Phi}&=&e^{2\tilde{\Phi}}\, U^{-1}\cr
H &=& dx^0\wedge dx^1 \wedge dU^{-1}\qquad\qquad ,
\eea
where the function $U(r,\rho)$ is given in (\ref{U}).
\section{Conclusions}
\cleqn

In this letter we have considered  the (super) gravity field equations, corresponding to
the low energy dynamics of non-critical string theory in dimensions $D=d+2<10$,
and focused on the bosonic solution that describes the non-critical string, the
analogue of the $F1$ solution in the critical case.
The backgrounds obtained, take into account the full back-reaction due to the presence of
a fundamental non-critical string in the vacuum defined by the direct product of flat $d$
dimensional Minkowski space-time and a two dimensional euclidean theory, the cigar.
They are determined by a particular Green function $U$, which replaces the usual
$\; 1+ \left(\frac{a}{r}\right)^{d-2}\;$ of $d$ dimensional euclidean space, that we have
explicitly constructed.
This solution should be supersymmetric for even $d$; however, in absence of the
complete (including fermions) supergravity theory, the analysis of SUSY remains no clear
to us in this framework.
It is worth to note that the solution refers to the presence of strings in a definite
vacuum that solves the equations (\ref{eqnmot}) without sources; we could choose another one,
for example, replacing the cigar by the CHS background
$\Re_\phi \times SU(2)_k$ \cite{chs}, which describes the near horizon geometry of $k$
$NS5$-branes
\footnote{
For a recent study of intersecting branes in a linear dilaton background, see \cite{cgo}.
}; in that case, the ansatz (\ref{ansatz}) must be generalized, but it is almost certain
the existence of a fundamental string solution like the one found here.

\begin{figure}[ht]
\centering

\includegraphics[clip,width=350pt]{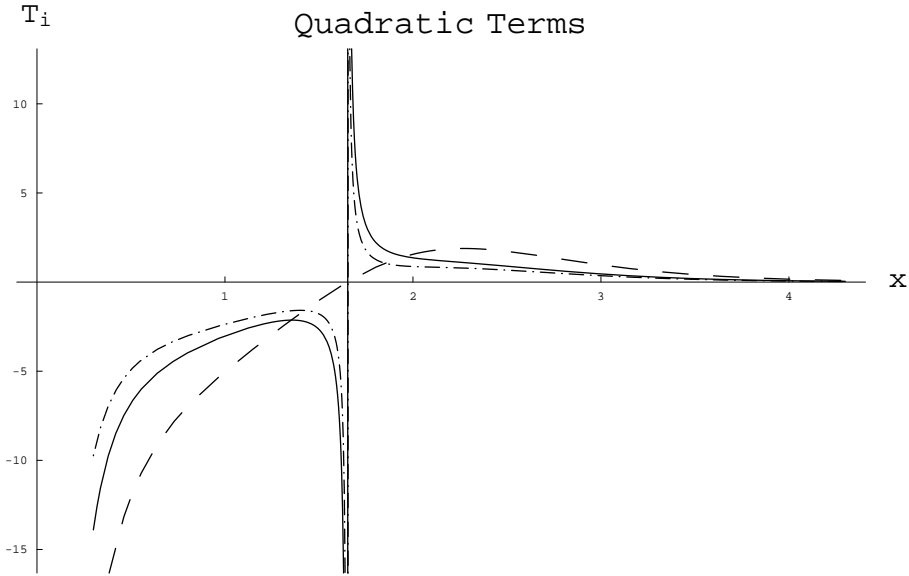}
\caption{Ratios between the correction terms and the $\alpha'R$
term, as a function of $x=\tilde{\rho}$, for $k=2/3$ and $\beta=10$.
$T_1=\alpha'^2R^{mnls}R_{mnls}/\alpha'R$\,\,(solid line),
$T_2=\alpha'^2R^{mn}R_{mn}/ \alpha'R$\,\,(dashed-dotted line),
$T_3=\alpha'^2 R^2/\alpha'R$\,\,(dashed line).
}
\label{grafico1}
\end{figure}

As remarked in the Introduction, an important fact to be addressed is the validity of
the background found.
Let us take for definiteness the $d=2$ case, where we have obtained an explicit form
for the function $U(\tilde\rho)$,
\be
U(\tilde\rho)=1 + \beta\; \ln \left(1+{1 \over \sinh^2\tilde{\rho}} \right)
\label{Ud2}
\ee
where $\beta= \frac{|Q|\, e^{2\tilde\phi_0}}{4\pi}$ is a numerical constant.
This solution is strictly valid far away from the tip of the cigar, as it is suggested
by the computation of potential second order corrections in the curvature
\footnote{
In ten dimensions, it is known that a quadratic Gauss-Bonnet type contribution
is present in heterotic theories, while in type II theories the first corrections
are quartic \cite{zgw}.
In non critical supergravities however, the quadratic corrections should not be
necessarily zero.
}.
These three terms are shown in Fig. \ref{grafico1}.
Even though, strictly speaking, we should compare terms in the equations of motion rather
in the action, it does not matter in the region of large $\tilde\rho$, where all the
terms goes asymptotically to zero
\footnote{
More precisely, the leading term $\alpha' R$ goes to zero like $e^{-2\tilde\rho}$,
and any of the three leading higher order corrections, like $e^{-4\tilde\rho}$.
}.
However it could do at finite $\tilde\rho$; the divergence observed in $T_1$ and $T_2$
is trivial, due to a zero in the curvature ($\sim T_3$), and it is not necessary present
when considering the equation of motion.
In fact, we conjecture that the solution remains valid not only in a region far away
from the tip of the cigar, but also in an intermediate region $\tilde\rho \sim 1$,
as it is the case for the typical extremal and black-brane solutions in critical dimension
\cite{hs}.
The problem of the higher order corrections becomes crucial when we look at a
``near horizon limit", i.e. when we go to the tip $\tilde\rho=0$.
It is in this case when, presumably from $D$-brane solutions, $Ads$ type solutions like
the ones investigated in \cite{km}, \cite{ks1}, emerge.
Constant curvature terms of order unity (instead of being controlled by a parameter like the
t'Hooft coupling) could invalidate the supergravity approximation as a framework
to study field theories via holography.
Nevertheless, it is widely conjectured that higher order corrections do not
affect the geometry of such backgrounds \cite{km}, \cite{ks1} \cite{ks2}.
We would like to remark however, that all these facts are not new; with exception of the
$D3$-brane (and the $NS5$-brane) which is regular at the horizon, all the known brane
solutions are valid in a bounded region of the space \cite{imsy}.

As a final remark, we could guess, from the presence of electric-magnetic duality of the
$B$-field equations on a fixed geometry, the existence of a ``$NS (D-5)$-brane" solution,
the analogue of the $NS5$-brane of CST, magnetically charged under $B$.
In relation to this point, we note that the usual ten dimensional S-duality transformation
that interchanges the $D5$-brane with the $NS5$-brane, and allows to construct the last one
from the knowledge of the $D5$-brane solution, is not present in NCST.
Subjects like the study of magnetic Neveu-Schwarz branes, as well as solutions of
$Dp$-branes
\footnote{
The construction of $D$-branes in non-critical string theories, as boundary states in
the bulk closed string theory, is addressed in some recent work, see i.e. references
\cite{fnp} , \cite{amt}.
}
, are certainly among the future lines of research, and will be presented elsewhere
\cite{lsprep}.

\section{Acknowledgements}
We would like to thank Carmen Nu\~nez and Fidel Schaposnik for useful discussions, and
specially Gast\'on Giribet for a careful reading of the manuscript and valuable suggestions.

\appendix

\section{Appendix}

We briefly review here the standard construction of Green functions, as applied to the
problem found in the paper.

Let us consider a metric space $(M,g)$, direct product of two ones, $(M_1,g_1)$
(with generic coordinates $x$) and $(M_2,g_2)$ (with coordinates $y$),
i.e. $M =M_1 \times M_2$, with metric $g=g_1 + g_2$.
Let $\hat{A}_1 , \hat{A}_2$ be two linear operators in the space of functions on
$M_1$ and $M_2$ respectively, and consider the differential equation,
\be
\hat{A}\,G(x,y;x_0 ,y_0) \equiv \left( \hat{A}_1+\hat{A}_2\right)G(x,y;x_0 ,y_0)
=\delta_{g_1}(x-x_0)\;\delta_{g_2}(y-y_0)\label{ecfgreen}
\ee
The function $G(x,y;x_0 ,y_0)$, solution to this equation, is the Green function of the
operator $\hat{A}$ w.r.t. the metric $g$.

Let us assume that there exist sets of eigen-functions $\{ u_m\}$ and $\{ v_i\}$,
with eigen-values $\{ \lambda_m\}$ and $\{ \mu_i\}$
\footnote{
For simplicity, we consider the no existence of zero modes.
We note, however, that in principle, for the existence of (\ref{fgreen}), is needed
that just $\hat A$ has no zero modes.
}
,
\bea
\hat{A}_1\, u_m(x)&=&\lambda_m\; u_m (x)\cr
\hat{A}_2\, v_i(y)&=& \mu_i\; v_i(y)\;\;\;\;,
\eea
where the labels ``m" and ``i" run over sets that can be discrete, continuous or both.
Let us further suppose that each set is orthonormal and complete in the sense,
\bea
\int_{M_1}\;\epsilon_{g_1}\; u^*_m\; u_n &=& \delta_{mn}\;\;\;\;,\;\;\;\;
\delta_{g_1}(x - x_0)=\sum_m\; u^*_m(x_0)\; u_m(x)\cr
\int_{M_2}\;\epsilon_{g_2}\; v^*_i\; v_j &=& \delta_{ij}\;\;\;\;\;\;,\;\;\;\;
\delta_{g_2}(y - y_0)=\sum_i\; v^*_i(y_0)\; v_i(y)
\eea
Then, the Green functions for the operators $\hat{A}_1\, ,\hat{A}_2\,$ can be written
\bea
G^{(1)}(x,x_0)&\equiv& \sum_m\; {u^*_m(x_0)\; u_m(x)\over \lambda_m}\;\;\;\;,\;\;\;\;
\hat{A}_1\, G^{(1)}(x,x_0)= \delta_{g_1}(x- x_0)\cr
G^{(2)}(y,y_0)&\equiv& \sum_i\; {v^*_i(y_0)\; v_i(y)\over \mu_i}\;\;\;\;\;\;\;\;,\;\;\;\;
\hat{A}_2\, G^{(2)}(y,y_0)= \delta_{g_2}(y- y_0)
\eea
Now, let us define,
\bea
G(x,y;x_0,y_0)&\equiv&\sum_{m,i}{u^*_m(x_0)\; v^*_i(y_0)\; u_m(x)\; v_i(y)\over
\lambda_m+\mu_i}\label{fgreen}
\eea
It is straightforward to prove that
\be
(\hat{A}_1+\hat{A}_2)\, G(x,y; x_0,y_0)=\delta_{g_1}(x - x_0)\; \delta_{g_2}(y-y_0)
\ee
That is, (\ref{fgreen}) is, according to (\ref{ecfgreen}), the Green function for the operator
$\hat A\equiv \hat{A}_1+\hat{A}_2$.

Going to the case of interest of Section $3$, let us first identify
$\hat{A}_1\equiv \Delta$, the laplacian in flat $\Re^{d-2}$.
A complete set of eigen-functions, orthonormal w.r.t. the flat metric in the $\delta$-function
sense, is certainly
\be
u_{\vec p }(\vec y) = {e^{i\vec p \cdot \vec y }\over (2\pi)^{d-2\over 2}}\qquad,\qquad
\int_{\Re^{d-2}}\, d^{d-2} \vec y\;u_{{\vec p}'}^*(\vec y)\; u_{\vec p}(\vec y) =
\delta^{d-2}({\vec p}'  - \vec p)\label{u}
\ee
with eigenvalues $\lambda_{\vec p}={\vec p}^2\;,\;\vec p\in\Re^{d-2}$.

On the other hand, let us take $\hat{A}_2\equiv \tilde L_0$, as defined in (\ref{L0});
standard manipulations yield the complete set of eigen-functions,
\footnote{
They are essentially the Jacobi functions
$\;{\cal P}^{-\frac{1}{2} +i\, s}_{-\frac{m}{2}\,\frac{m}{2}}\;$, see \cite{dvv} and
references therein.
}
\bea
v_{s m} (\vec z) &=& a_{sm}\; \left(\frac{\rho}{r_0}\right)^{|m|}\;
F\left(\frac{|m| + 1}{2} + i\,s, \frac{|m| + 1}{2} -i\,s; |m| +1;
-\left(\frac{\rho}{r_0}\right)^2\right)\; \frac{e^{i\,m\,\theta}}{\sqrt{2\,\pi}}\cr
a_{sm} &=& \left(\frac{4}{\pi\,r_0{}^2}\; \rho_m(s)\right)^\frac{1}{2}\;
\frac{i^{-|m|}}{|m|!}\;
\frac{\Gamma(\frac{|m| + 1}{2} + i\,s)}{\Gamma(\frac{-|m| + 1}{2} + i\,s)}\cr
&=&
\left(\frac{4}{r_0{}^2}\;s\;Re\tanh\left(\pi\,(s+i\,\frac{m}{2})\right)\right)^\frac{1}{2}\;
\frac{i^{-|m|}}{|m|!}\;
\frac{\Gamma( \frac{|m| + 1}{2} + i\,s)}{ \Gamma(\frac{-|m| + 1}{2} + i\,s)}
\eea
where $F(a,b;c;z)$ is the hypergeometric function, and $\rho_m(s)$ is the so called
Plancherel measure.
They are orthonormal w.r.t. the flat metric $\;\delta = d\vec z\cdot d\vec z = d^2\rho +
\rho^2\,d^2\theta\,$, introduced in (\ref{flat2dmetric}),
\be
\int_{\Re^2}\; d^2 \vec z\; v_{s' m'}^* (\vec z)\; v_{s m} (\vec z) = \delta(s'
-s)\;\delta_{m' m}\;\;\;\;,
\ee
and their $\tilde L_0$-eigenvalues are given by
\be
\mu_{sm} = \frac{1}{r_0{}^2}\,\left( 4\, s^2 + m^2 +1 \right)
\qquad,\qquad s\in\Re^+\;\;,\;\; m\in\bz
\ee
From these results we can construct the Green function (\ref{fgreen}), leading
to the functions $U$ in (\ref{U}), (\ref{Ud2}), solutions of the equation (\ref{eqnUbis}).


\end{document}